\documentclass[11pt,a4paper]{article}
\renewcommand{\epsilon}{\varepsilon}
\begin{document}
\title{On the mode structure of imperfect fluids}
\author{Winfried Zimdahl\\
\ \\
N\'ucleo COSMO-UFES \& Departamento de F\'isica\\  Universidade Federal do Esp\'irito Santo (UFES)\\
 Av. Fernando Ferrari s/n CEP 29.075-910, Vit\'oria, ES, Brazil}

\maketitle


\begin{abstract}
This paper tries to present a simple picture of several aspects of the mode structure in relativistic non-equilibrium thermodynamics.
Its pedagogical focus is on the relation between long-wavelength perturbation modes of the causal M\"{u}ller-Israel-Stewart (MIS) theory and those of the traditional Eckart theory.
Principally, this issue was clarified in a series of papers by Hiscock and Lindblom (see \cite{HiLi,HiLi85,HiLi2}).
Here, I put together some essential features of this topic which do not require the entire formalism of the complete theory.
\end{abstract}
\section{Introduction}

Traditionally, thermodynamical processes out of equilibrium are described
by the theories of Eckart \cite{Eckart:1940te} and Landau and Lifshitz \cite{LL58}.
With the works by M\"{u}ller \cite{M67}, Israel \cite{I76}, Israel and
Stewart \cite{ISAnn79,IS79}, Pav\'{o}n, Jou, and Casas-V\'{a}zquez \cite{Pavon82}, Hiscock
and Lindblom \cite{HiLi,HiLi85,HiLi2} it became clear, however, that the traditional
theories suffer from serious drawbacks concerning causality
and stability. These difficulties could be traced back to
their restriction to first-order deviations from thermodynamical equilibrium. If
one includes second-order deviations as well, the corresponding
problems disappear. By now, it is generally agreed that
any analysis of dissipative phenomena in relativity should be
based on the theories by M\"{u}ller, Israel, and Stewart (MIS),
although, in specific cases, the latter might reproduce results
of the Eckart theory  \cite{HiLi2}. Cosmological implications of
second-order theories, also called causal thermodynamics, were first considered by Belinskii
et al. \cite{Belinski79}.
Of particular interest in this context have been bulk-viscous cosmological models \cite{diego,Luis,Jou,Roy,ZPM,Mak,Ich,OliverJulio}.

While conceptually the idea of including second-order deviations from equilibrium
is quite clear, the detailed implementation requires a rather extended formalism.
Therefore it might be desirable to find a simplified but nevertheless exact (within a
certain range) account of essential features of causal thermodynamics
which focusses in detail on the points where it differs from Eckart-type theories.
One of the shortcomings of the latter theories is their prediction of instabilities
of perturbation modes on very short time scales.
Our aim here is to clarify the origin of this different behavior for the case of
long-wavelength modes in flat space-time.

We start our analysis by recalling basic relations of imperfect fluid dynamics in Sec.~\ref{Imperfect}.
This implies the structure of the energy-momentum tensor and that of the particle-flow vector. It follows a discussion of the conservation laws for particle number, energy and momentum, which includes the constitutive relations for the thermodynamic fluxes. Through these relations the difference between MIS and Eckart-type theories becomes manifest.
Up to this point the formalism is kept general.
In Sec.~\ref{Linearization} we specify the general dynamics to linear perturbations about a fluid at rest in Minkowski-space. This excludes all gravitational degrees of freedom. But since fluid-dynamical scales in most applications are smaller than gravitational scales, e.g., the Hubble scale in cosmology, the restriction to a flat background may nevertheless capture astrophysically and cosmologically relevant situations.
Considering plane-wave solutions, the linearized conservation equations are then boiled down in Sec.~\ref{Modes} to a system of coupled algebraic equations in terms of perturbations of the number density and the temperature. This system serves as a starting point to investigate the long-wavelength mode structure, the main topic of this paper. We identify the origin of instabilities in Eckart's theory and point out how the MIS theory avoids such unphysical behavior.
We emphasize that the propagation of acoustic modes including their damping through viscosity and heat conductivity, originally obtained within Eckart's theory, remains exactly valid in causal thermodynamics as well.

\section{Imperfect fluids}
\label{Imperfect}
\subsection{Energy-momentum tensor and particle flow vector}
The energy-momentum tensor of an imperfect fluid is generally given by (greek indices run over 0,1,2,3)
\begin{equation}
T ^{\alpha\beta} = T ^{\alpha\beta}_{\left(0 \right)} + \pi h ^{\alpha\beta}
+ \pi^{\alpha\beta} + q^{\alpha}u ^{\beta} + q ^{\beta}u ^{\alpha}
\label{mT}
\end{equation}
with
\begin{equation}
T ^{\alpha\beta}_{\left(0 \right)} = \rho  u ^{\alpha}u ^{\beta} + p h ^{\alpha\beta}
\label{mT0}
\end{equation}
and
\begin{equation}
\pi ^{\alpha\beta}u _{\beta} = q ^{\alpha}u _{\alpha} =
\pi ^{\alpha}_{\alpha} =  h ^{\alpha\beta}u _{\alpha} = 0,
\ \ \ \ \ \ u ^{\alpha}u _{\alpha} = -1.
\label{mort}
\end{equation}
Here, $\rho$ is the energy density of a fiducial thermodynamical equilibrium state, represented by the part $T ^{\alpha\beta}_{\left(0\right)}$ of the total energy-momentum tensor,
$p$ is the corresponding equilibrium pressure, $u ^{\alpha}$ is the fluid four-velocity in the Eckart frame and $h ^{\alpha\beta} = g ^{\alpha\beta} + u ^{\alpha}u ^{\beta}$
is the spatial projection tensor.
The speed of light has been normalized to unity.
The quantity $\pi $ denotes that part of the scalar pressure which is
connected with entropy production, $- \pi ^{\alpha\beta}$ is the anisotropic
stress tensor and
$q ^{\alpha}$ is the heat flux vector.
Within the Eckart frame the particle number flow vector $N ^{\alpha}$ is
given by
\begin{equation}
N ^{\alpha} = n u ^{\alpha},
\label{mN}
\end{equation}
where $n$ is the particle number density.

\subsection{Conservation equations}
The basic set of hydrodynamical equations follows from the conservation laws
$N ^{\alpha}_{;\alpha} = 0$ and $T ^{\alpha\beta}_{\  ;\beta}= 0$.
This implies particle number conservation
\begin{equation}
\dot{n} + \Theta n = 0,
\label{mdn}
\end{equation}
where $\Theta \equiv  u ^{\alpha}_{;\alpha}$ is the fluid expansion scalar and
$\dot{n} \equiv  n _{,\alpha}u ^{\alpha}$, as well as
the energy conservation
\begin{equation}
\dot{\rho } + \Theta \left(\rho + p + \pi \right)
+ \nabla _{\alpha}q ^{\alpha} + 2 \dot{u}_{\alpha}q ^{\alpha}
+ \sigma _{\alpha\beta}\pi ^{\alpha\beta} = 0
\label{mdr}
\end{equation}
and the momentum conservation
\begin{equation}
\left(\rho + p + \pi  \right)\dot{u}_{\alpha}
+ \nabla  _{\alpha}\left(p + \pi  \right)
+ \nabla  ^{\beta}\pi _{\alpha\beta}
+ \dot{u}^{\beta}\pi _{\alpha\beta}
+ h ^{\beta}_{\alpha}\dot{q}_{\beta}
+ \left[\omega _{\alpha\beta} + \sigma _{\alpha\beta} + \frac{4}{3}\Theta h _{\alpha\beta} \right]
q ^{\beta} = 0,
\label{mconsm}
\end{equation}
where $\nabla _{\alpha}q ^{\alpha} \equiv  h _{\alpha}^{\beta}q ^{\alpha}_{;\beta}$ etc.
The quantity $\sigma _{\alpha\beta}$ is the shear tensor
\begin{equation}
\sigma _{\alpha\beta} = \frac{1}{2}\left(\nabla _{\alpha}u _{\beta}
+ \nabla  _{\beta}u _{\alpha} - \frac{2}{3}h _{\alpha\beta}\Theta \right)
\label{msigma}
\end{equation}
and $\omega_{\alpha\beta}$ is the vorticity tensor
\begin{equation}
\omega _{\alpha\beta} = \frac{1}{2}\left(\nabla _{\alpha}u _{\beta}
- \nabla  _{\beta}u _{\alpha} \right).
\label{momega}
\end{equation}
The energy conservation (\ref{mdr}) is the result of the projection $u _{\alpha}T^{\alpha\beta}_{\  ;\beta}= 0$ of $T^{\alpha\beta}_{\  ;\beta}= 0$ while the momentum conservation (\ref{mconsm}) follows from the orthogonal projection
$h_{\alpha\mu}T^{\mu\beta}_{\  ;\beta}= 0$.
We assume equations of state of the general form
\begin{equation}
p = p \left(n,T \right),\ \ \ \ \ \ \ \ \
\rho = \rho \left(n,T \right),
\label{meos}
\end{equation}
i.e., we will use the particle number density $n$ and the temperature $T$ as independent thermodynamical variables.
Within the MIS  theory the thermodynamic ``fluxes'' $\pi$, $q _{\alpha}$ and
$\pi _{\alpha\beta}$ obey the following evolution equations, in which, for simplicity, we have neglected the couplings between heat flux and viscous pressures \cite{HiLi}:
\begin{equation}
\pi = - \zeta \left[\Theta + \beta _{0}\dot{\pi }
+ \frac{\pi }{2}T \left(\frac{\beta _{0}}{T}u^{\gamma} \right)_{;\gamma} \right],
\label{anspi}
\end{equation}
\begin{equation}
q ^{\alpha} = - \lambda h ^{\alpha\beta}
\left[\nabla  _{\beta}T + T \dot{u}_{\beta} + T\beta _{1}\dot{q}_{\beta}
+ \frac{T ^{2}}{2}q _{\beta}
\left(\frac{\beta _{1}}{T}u ^{\gamma} \right)_{;\gamma}
\right],
\label{ansq}
\end{equation}
and
\begin{equation}
\pi ^{\alpha\beta} = - 2 \eta h ^{\alpha\mu}h ^{\beta\nu} \left[\sigma _{\mu\nu}
+ \beta _{2}\dot{\pi }_{\mu\nu}
+ \frac{\pi _{\mu\nu}}{2}\left(\frac{\beta _{2}}{T}u ^{\gamma} \right)_{;\gamma}\right].
\label{anspit}
\end{equation}
The symbols $\zeta$, $\lambda$ and $\eta$ denote the (positive) coefficients of bulk viscosity, heat conductivity and shear viscosity, respectively.
Additionally, the second-order theory is characterized by the (positive) coefficients $\beta _{0}$, $\beta _{1}$ and $\beta _{2}$.
These coefficients appear in terms with time derivatives of $\pi$, $q _{\alpha}$ and
$\pi _{\alpha\beta}$, respectively.
The appearance of time derivatives of $\pi$, $q _{\alpha}$ and
$\pi _{\alpha\beta}$ makes the relations (\ref{anspi}) - (\ref{anspit}) evolution equations which ensure that the entropy production is non-negative \cite{HiLi}.
The Eckart case corresponds to $\beta _{0}=\beta _{1}=\beta _{2}=0$, which cancels these time derivatives  and reduces the set (\ref{anspi}) - (\ref{anspit}) to algebraic equations.

\section{Linearization procedure}
\label{Linearization}
\subsection{Background}
In a first step we specify the relations of the previous chapter to a homogeneous, isotropic perfect fluid at rest in Minkowski space.
The corresponding quantities are denoted by an overbar.
We have
\begin{equation}
\bar{u}^{0} = 1 \ ,
\bar{u}_{0} = -1 \ ,
\bar{u}^{a} = \bar{u}_{a}= 0,
\label{mzero}
\end{equation}
(latin indices run over 1,2,3) and
\begin{equation}
\bar{\Theta }=\nabla  _{\alpha}\bar{T}=\bar{\sigma}_{\alpha\beta}
= \bar{\dot{u}}_{\alpha} =
\bar{\pi}= \bar{\pi}^{\alpha\beta}=\bar{q}^{\alpha}=0.
\label{mzero2}
\end{equation}
It follows that
\begin{equation}
\bar{n}= {\rm const},\
\bar{T}={\rm const},\
\bar{\rho }= {\rm const},\
\bar{p}= {\rm const}.
\label{mconst}
\end{equation}
\subsection{First-order perturbations}
Now we consider perturbations of all the thermodynamic quantities, denoted by a hat symbol:
\begin{equation}
n = \bar{n}+\hat{n}, \quad
T = \bar{T}+\hat{T},\quad
\rho = \bar{\rho }+\hat{\rho },\quad
p=\bar{p}+\hat{p}.
\label{mpertnT}
\end{equation}
``First order" here means always linear in $\hat{n}$, $\hat{T}$, $\hat{\rho}$ and $\hat{p}$.
Perturbing also the relation
$\eta _{\mu\nu}u ^{\mu}u ^{\nu}=-1$ yields at linear order,
\begin{equation}
\eta _{\mu\nu}\hat{u}^{\mu}\bar{u}^{\nu}=0
\quad\Rightarrow\quad
\hat{u}^{0}=\hat{u}_{0}=0.
\label{mhu0}
\end{equation}
For the perturbed spatial components of the four-velocity we have $\hat{u}^{m}=\hat{u}_{m}$  and the first-order expansion scalar is
\begin{equation}
\hat{\Theta}=\hat{u}^{m}_{,m}.
\label{mhT}
\end{equation}
For the first-order perturbations of the acceleration we find
\begin{equation}
\left(\dot{u}_{\mu}\right)\hat{} = \hat{u}_{\mu,\nu}u ^{\nu}
+ u_{\mu,\nu}\hat{u}^{\nu}
\quad\Rightarrow\quad
\left(\dot{u}_{0}\right)\hat{}=0,
\mbox{\ \ }
\left(\dot{u}_{m,m} \right)\hat{} = \dot{\hat{\Theta}}.
\label{mdhT}
\end{equation}
Furthermore, we have at first order
\begin{equation}
h _{0}^{\alpha}\dot{q}_{\alpha} = 0,
\mbox{\ \ \ \ }
h _{m}^{\alpha}\dot{q}_{\alpha} = \dot{q}_{m}.
\label{mdq}
\end{equation}
The terms
$\dot{u}_{\alpha}q ^{\alpha}$,
$\sigma _{\alpha\beta}\pi ^{\alpha\beta}$,
$\pi \dot{u}_{\mu}$,
$\dot{u}^{\beta}\pi _{\alpha\beta}$ and
$\left[\omega _{\alpha\beta}+\sigma _{\alpha\beta}+\frac{4}{3}\Theta h _{\alpha\beta} \right]q ^{\beta}$ are of second order and will be omitted in the following.

\subsection{First-order conservation equations}
The linearized set of equations becomes (now we omit the overbars for the background variables)
\begin{equation}
\dot{\hat{n}} + n \hat{\Theta }= 0,
\label{mdhn}
\end{equation}
\begin{equation}
\dot{\hat{\rho }} + \left(\rho + p \right)\hat{\Theta }
+ q _{a,a} = 0,
\label{mdhr}
\end{equation}
and
\begin{equation}
\left(\rho + p \right)\left(\dot{u}_{m} \right)\hat{}
+ \hat{p}_{,m}
+ \pi _{,m} + \pi _{m a,a}
+ \dot{q}_{m} = 0.
\label{mmomcons}
\end{equation}
Taking the spatial divergence of Eq. (\ref{mmomcons}) and applying the last of the relations (\ref{mdhT}), we get
\begin{equation}
\left(\rho + p \right)\dot{\hat{\Theta}} + \Delta \hat{p} + \Delta \pi
+ \pi _{m a,m a}
+ \dot{q}_{m,m} = 0,
\label{mdivmom}
\end{equation}
where $\Delta $ is the three-dimensional Laplacian.
By using (\ref{mdivmom}) together with (\ref{mdhn}) and (\ref{mdhr}) we shall obtain the longitudinal modes.
The transverse modes will be found by taking the spatial rotation of (\ref{mmomcons}):
\begin{equation}
2 \left(\rho + p \right)\dot{w}_{a}
+ \epsilon _{amn}\left(\pi _{mb,bn}
+ \dot{q}_{m,n}\right) = 0,
\label{mtransv}
\end{equation}
where
\begin{equation}
w _{a} \equiv  \frac{1}{2}\epsilon _{amn}\hat{u}_{m,n}
\label{mw}
\end{equation}
and $\epsilon _{amn}$ is the three-dimensional Levi-Civita symbol.

These are the basic equations from which we shall derive the explicit mode structure in the  following section.

\section{Mode structure}
\label{Modes}
\subsection{Longitudinal modes}
For the dissipative terms in (\ref{mdhr}) and (\ref{mdivmom}) we find up to linear order
\begin{equation}
\Delta \pi = - \zeta \left[\Delta \hat{\Theta}
+ \beta _{0}\Delta \dot{\pi}\right],
\label{mdpi}
\end{equation}
\begin{equation}
q _{a,a} = - \lambda T
\left[\frac{\Delta T}{T} + \dot{\hat{\Theta}}
+ \beta _{1}\dot{q}_{a,a} \right],
\label{mdivq}
\end{equation}
and
\begin{equation}
\pi _{ma,ma} = - 2 \eta
\left[\frac{2}{3}\Delta \hat{\Theta} + \beta _{2}\dot{\pi}_{ma,ma} \right],
\label{mdtpi}
\end{equation}
respectively.
The divergence terms in (\ref{anspi})-(\ref{anspit}) are of second order and do not contribute in (\ref{mdpi})-(\ref{mdtpi}).
In (\ref{mdtpi}) we have used that at first order
\begin{equation}
\sigma _{00}=\sigma _{0 b} = \sigma _{b0} = 0 \qquad \mathrm{and} \qquad
\sigma _{ab,ab} =
\frac{2}{3}\Delta \hat{\Theta}.
\label{msig}
\end{equation}
Upon using the equations of state (\ref{meos}) we may replace
$\dot{\hat{\rho}}$ by
\begin{equation}
\dot{\hat{\rho}} = \left(\rho + p - T \frac{\partial{p}}{\partial{T}} \right)
\frac{\dot{\hat{n}}}{n}
+ \frac{\partial{\rho}}{\partial{T}}\dot{\hat{T}}
\label{mdhrnT}
\end{equation}
and $\Delta \hat{p}$ by
\begin{equation}
\Delta \hat{p} = \frac{\partial{p}}{\partial{n}}\Delta \hat{n}
+ \frac{\partial{p}}{\partial{T}}\Delta \hat{T}.
\label{mdeltap}
\end{equation}
To obtain (\ref{mdhrnT}) we have used the thermodynamic relation
\begin{equation}\label{relthermo}
\frac{\partial \rho}{\partial n} = \frac{\rho + p}{n} - \frac{T}{n}\frac{\partial p}{\partial T},
\end{equation}
which guarantees that the entropy $s$ per particle, defined by
\begin{equation}\label{}
s = \frac{\rho + p}{nT} - \frac{\mu}{T},
\end{equation}
where $\mu$ is the chemical potential,
is a state function.
Eq.~(\ref{relthermo}) then follows from the requirement that the second derivatives of $s$  with respect to the basic thermodynamical variables be interchangeable:
\begin{equation}\label{2nd}
\frac{\partial^{2}s}{\partial n \partial T} = \frac{\partial^{2}s}{\partial T \partial n}.
\end{equation}
The set of independent variables consists of $\hat{n}$, $\hat{T}$, and
$\hat{\Theta}$ as well as of $\pi$, $q _{a}$, and
$\pi _{ma}$.
Now we look for plane-wave solutions
\begin{equation}
\hat{n}\ , \hat{T}\ , \hat{\Theta }, ...\propto
\exp{\left[i \left(\omega t - k _{a}x ^{a}\right) \right]}.
\label{mpw}
\end{equation}
From the perturbed particle number conservation equation (\ref{mdhn}) we find
\begin{equation}
\hat{\Theta } = - i \omega \frac{\hat{n}}{n}.
\label{mhT}
\end{equation}
Since $\hat{\Theta }$ may always be eliminated with the help of the last relation, we will end up with a system for
$\hat{n}$ and $\hat{T}$.
The energy conservation (\ref{mdhr}) with (\ref{mdhn}), (\ref{mdhrnT}) and (\ref{mhT}) then becomes
\begin{equation}
-i \omega T \frac{\partial{p}}{\partial{T}}\frac{\hat{n}}{n}
+ i \omega \hat{T} \frac{\partial{\rho }}{\partial{T}}
- i k _{a}q _{a} = 0,
\label{mbalen}
\end{equation}
while the  momentum conservation (\ref{mdivmom}) with (\ref{mdeltap}), (\ref{mhT})  and
$\dot{\hat{\Theta }} \rightarrow i \omega \hat{\Theta }
= \omega ^{2}\hat{n}/n$ is
\begin{equation}
\left[\left(\rho + p \right)\omega ^{2}
- k ^{2} n \frac{\partial{p}}{\partial{n}} \right]
\frac{\hat{n}}{n}
- k ^{2}T \frac{\partial{p}}{\partial{T}}\frac{\hat{T}}{T}
- k ^{2}\pi - k _{m}k _{a}\pi _{m a}
- i \omega ik _{m}q _{m} = 0,
\label{mbalm}
\end{equation}
where $k ^{2}\equiv  k _{a}k ^{a}$.
For the dissipative quantities we find from (\ref{mdpi})-(\ref{mdtpi})
\begin{equation}
- k ^{2}\pi
= -i \omega \frac{\zeta k ^{2}}{1 + i \omega \zeta \beta _{0}}
\frac{\hat{n}}{n},
\label{mk2pi}
\end{equation}
\begin{equation}
- i k _{a}q _{a}
= - \frac{\lambda T}
{1 + i \omega \lambda T \beta _{1}}
\left[- k ^{2}\frac{\hat{T}}{T} + \omega ^{2} \frac{\hat{n}}{n} \right],
\label{mkq}
\end{equation}
and
\begin{equation}
- k_{a}k _{b}\pi _{ab}
= - \frac{4}{3}i \omega k ^{2}\frac{\eta }{1 + 2i \omega  \eta \beta _{2}}
\frac{\hat{n}}{n}.
\label{mk2tpi}
\end{equation}
With the help of the definitions
\begin{eqnarray}
\zeta _{\beta} & \equiv  & \frac{\zeta}{1 + i \omega \zeta \beta _{0}},
\label{zbeta}\\
\lambda _{\beta} & \equiv  & \frac{\lambda}
{1 + i \omega \lambda T\beta _{1}},
\label{lbeta}\\
\eta  _{\beta} & \equiv  & \frac{\eta}{1 + 2 i \omega \eta \beta _{2}},
\label{etabeta}
\end{eqnarray}
we may write
\begin{equation}
- k ^{2}\pi =  - i \omega k ^{2}\zeta _{\beta}\frac{\hat{n}}{n},
\label{mk2pib}
\end{equation}
\begin{equation}
- i k _{a}q _{a}  = - \lambda _{\beta}T
\left[- k ^{2}\frac{\hat{T}}{T} + \omega ^{2} \frac{\hat{n}}{n}\right],
\label{mkqb}
\end{equation}
and
\begin{equation}
- k _{a}k _{b}\pi _{ab}
=  - \frac{4}{3}i \omega k ^{2}\eta _{\beta}\frac{\hat{n}}{n},
\label{mk2tpib}
\end{equation}
respectively.
The definitions (\ref{zbeta})-(\ref{etabeta}) were chosen such that
formally the structures (\ref{mk2pib})-(\ref{mk2tpib}) are those of the Eckart theory.  This will allow us to treat the first- and second-order theories in parallel.

Combining the conservation equations (\ref{mbalen}) and (\ref{mbalm}) with (\ref{mk2pib})-(\ref{mk2tpib}),
our system reduces to
\begin{equation}
\left[\omega T \frac{\partial{p}}{\partial{T}}
- i \omega ^{2}\lambda _{\beta } T \right]\frac{\hat{n}}{n}
+ \left[- \omega T \frac{\partial{\rho }}{\partial{T}}
+ i k ^{2}\lambda _{\beta} T \right]\frac{\hat{T}}{T} = 0
\label{msystnT}
\end{equation}
and
\begin{eqnarray}
\left[\omega ^{2}
- \frac{n}{\rho + p}\frac{\partial{p}}{\partial{n}} k ^{2}
- i \omega k ^{2}\left(\frac{\frac{4}{3}\eta _{\beta }
+ \zeta _{\beta }}{\rho + p}\right)
- i \omega ^{3}\frac{\lambda _{\beta }T}{\rho + p}
\right] \frac{\hat{n}}{n}&& \nonumber \\
\qquad\qquad\qquad + \left[- \frac{T}{\rho + p}\frac{\partial{p}}{\partial{T}}k ^{2}
+ i \omega k ^{2}\frac{\lambda _{\beta }T}{\rho + p} \right]
\frac{\hat{T}}{T}
&=& 0,
\label{msystnT2}
\end{eqnarray}
respectively.
The set of equations (\ref{msystnT}) and (\ref{msystnT2}) describes first-order perturbations within the MIS theory where the coupling between heat flux and viscous pressures was neglected.
The corresponding perturbations of the Eckart theory follow for
$\zeta  _{\beta } \rightarrow \zeta  $,
$\lambda _{\beta }\rightarrow \lambda $, and
$\eta _{\beta} \rightarrow \eta $.
Notice that $\zeta  _{\beta }$,
$\lambda _{\beta }$, and
$\eta _{\beta }$ depend on $\omega $.

The system (\ref{msystnT}) and (\ref{msystnT2}) provides us with the relation
\begin{eqnarray}
\left[\omega ^{2}
- \frac{n}{\rho + p}\frac{\partial{p}}{\partial{n}} k ^{2}
- i \omega k ^{2}\left(\frac{\frac{4}{3}\eta _{\beta }
+ \zeta _{\beta }}{\rho + p}\right)
- i \omega ^{3}\frac{\lambda _{\beta }T}{\rho + p}
\right]
\left[- \omega T \frac{\partial{\rho }}{\partial{T}}
+ i k ^{2}\lambda _{\beta }T \right]
&& \nonumber  \\
- \left[- \frac{T}{\rho + p}\frac{\partial{p}}{\partial{T}}k ^{2}
+ i \omega k ^{2}\frac{\lambda _{\beta }T}{\rho + p} \right]
\left[\omega T \frac{\partial{p}}{\partial{T}}
- i \omega ^{2}\lambda _{\beta }T \right]&& = 0.\nonumber\\
\label{msyst}
\end{eqnarray}
By multiplying this equation by
$-\left(T \partial \rho / \partial T \right)^{-1}$
and introducing the square of the sound velocity $c _{s}^{2}$ by
\begin{equation}
c _{s}^{2}  \equiv  \frac{n}{\rho + p}\frac{\partial{p}}{\partial{n}}
+ \frac{T}{\rho + p}\frac{\left(\frac{\partial{p}}{\partial{T}} \right)^{2}}
{\frac{\partial{\rho }}{\partial{T}}},
\label{mcs}
\end{equation}
we obtain the dispersion relation
\begin{eqnarray}
&&-i \omega ^{4}\frac{\lambda _{\beta }T}{\rho + p} + \omega ^{3}
- \omega k ^{2}c _{s}^{2}   - i \omega ^{2}k ^{2}
\frac{\frac{4}{3}\eta _{\beta }
+ \zeta _{\beta }}{\rho + p}
\nonumber\\
&&\qquad- i \omega ^{2}k ^{2}\lambda _{\beta }T
\left[\frac{1}{T \frac{\partial{\rho }}{\partial{T}}}
- \frac{2}{\rho + p}\frac{\frac{\partial{p}}{\partial{T}}}
{\frac{\partial{\rho }}{\partial{T}}} \right] + i k ^{4}\lambda _{\beta }\frac{n}{\rho + p}
\frac{\frac{\partial{p}}{\partial{n}}}{\frac{\partial{\rho }}{\partial{T}}} = 0.
\label{mdisp}
\end{eqnarray}
We have neglected here products of dissipative quantities.
It is obvious that the perfect-fluid limit is
\begin{equation}
\omega ^{2} = c _{s}^{2} k ^{2},
\mbox{\ \ \ \ \ \ \ \ \ \ \ \ \ \ \ }
\left(\zeta = \lambda = \eta = 0 \right).
\label{mo2id}
\end{equation}
With the help of the definitions (\ref{zbeta}), (\ref{lbeta}), and (\ref{etabeta}) for
$\zeta  _{\beta }$,
$\lambda _{\beta }$, and
$\eta _{\beta }$, respectively and neglecting higher-order terms in
$\zeta $,
$\lambda $, and
$\eta $, we may write (\ref{mdisp}) as
\begin{eqnarray}
&&-i \omega ^{4}\frac{\lambda T}{\rho + p}
+ \left(\omega ^{3}
- \omega k ^{2}c _{s}^{2}  \right)
\left[\left(1 + i \omega \zeta \beta _{0} \right)
\left(1 + i \omega \lambda T \beta _{1} \right)
\left(1 + 2i \omega \eta  \beta _{2} \right)
\right]\nonumber\\
&&\ - i \omega ^{2}k ^{2}
\frac{\frac{4}{3}\eta
+ \zeta }{\rho + p}
- i \omega ^{2}k ^{2}\lambda T
\left[\frac{1}{T \frac{\partial{\rho }}{\partial{T}}}
- \frac{2}{\rho + p}\frac{\frac{\partial{p}}{\partial{T}}}
{\frac{\partial{\rho }}{\partial{T}}} \right]
+ i k ^{4}\lambda\frac{n}{\rho + p}
\frac{\frac{\partial{p}}{\partial{n}}}{\frac{\partial{\rho }}{\partial{T}}} = 0.\nonumber\\
\label{mdisp2}
\end{eqnarray}
Let us now look at the long-wavelength limit $k \rightarrow 0$. We obtain
\begin{equation}
-i \omega ^{4}\frac{\lambda T}{\rho + p}
+ \omega ^{3}
\left[\left(1 + i \omega \zeta \beta _{0} \right)
\left(1 + i \omega \lambda T \beta _{1} \right)
\left(1 + 2i \omega \eta  \beta _{2} \right)
\right] = 0
\mbox{\ \ \ \ }
\left(k \rightarrow 0 \right).
\label{mdispk0}
\end{equation}
The last relation allows us to point out the different stability behavior of the MIS theory compared with Eckart's theory.
Besides of the always existing threefold solution $\omega = 0$ we have, in the Eckart case
$\beta _{0}=\beta _{1}=\beta _{2}=0$, the solution
\begin{equation}
i \omega _{L _{E}} = \frac{\rho + p}{\lambda T}.
\label{moE}
\end{equation}
As long as the right-hand side of (\ref{moE}) is positive, this means an imaginary frequency which, according to (\ref{mpw}),
describes an exponential instability on extremely short timescales which are much below any hydrodynamic scale \cite{HiLi85}.

It is obvious, how the situation changes in the MIS theory, where we find, besides of $\omega = 0$ (threefold)
\begin{eqnarray}
i \omega _{L _{0}} &=& - \frac{1}{\beta _{0}\zeta },
\label{mo0}\\
i \omega _{L _{1}} &=& - \frac{\rho + p}{\lambda T
\left[\beta _{1}\left(\rho + p \right) -1\right]},
\label{mo1}\\
i \omega _{L _{2}} &=& - \frac{1}{2\beta _{2}\eta },
\label{mo2}
\end{eqnarray}
instead of (\ref{moE}).
Since $\beta _{1}\left(\rho + p \right) > 1$ (see formula (134) in \cite{HiLi}), we conclude, that none of these modes is unstable.
While there is no counterpart of the modes (\ref{mo0}) and (\ref{mo2}) in the Eckart theory, the unstable mode (\ref{moE}) of the latter becomes stable according to (\ref{mo1}).
(For comparison: The modes (\ref{mo0})-(\ref{mo2}) are (37)-(39) in \cite{HiLi2}).

All the modes (\ref{moE})-(\ref{mo2}) describe perturbations which are far away from the perfect fluid behaviour (\ref{mo2id}).
In fact, in obtaining (\ref{mdispk0}) by formally putting $k=0$ in (\ref{mdisp2}), one considers the dissipative terms retained in (\ref{mdispk0}) to be of higher order than the perfect fluid contribution which leads to $c _{s}^{2}k ^{2}$ in (\ref{mo2id}).
These modes do not respect the requirement that dissipative terms should provide small correction to the perfect-fluid behavior.
The modes (\ref{mo0})-(\ref{mo2}) are strongly damped on time scales much smaller than any hydrodynamic time scale.

Next we study the dispersion relation for small but finite values of $k$. Up to linear order in the transport coefficients, Eq. (\ref{mdisp2}) is equivalent to
\begin{eqnarray}
&&i \omega ^{4} \left[\frac{\lambda T}{\rho + p}
\left(\beta _{1}\left(\rho + p \right) - 1 \right)
+ \beta _{0}\zeta + 2 \beta _{2}\eta \right]
+ \omega ^{3} + \omega k ^{2}c _{s}^{2}  \nonumber\\
&&\quad- \frac{i \omega ^{2}k ^{2}}{\rho + p}
\left[\left(\frac{4}{3}
+ 2 c _{s}^{2} \beta _{2}\left(\rho + p \right) \right)\eta
+ \left(1 + \beta _{0}\left(\rho + p \right) \right) \zeta \right.\nonumber\\
&&\qquad\qquad\qquad\left.
+ \lambda T \left(\beta _{1}c _{s}^{2} \left(\rho + p \right)
+ \frac{\rho + p}{T \frac{\partial{\rho }}{\partial{T}}}
- 2 \frac{\frac{\partial{p}}{\partial{T}}}
{\frac{\partial{\rho }}{\partial{T}}}\right)\right]
\nonumber\\
&&\quad+ i k ^{4}\lambda\frac{n}{\rho + p}
\frac{\frac{\partial{p}}{\partial{n}}}{\frac{\partial{\rho }}{\partial{T}}}  = 0.
\label{mo4}
\end{eqnarray}
We are now interested in the solutions $\omega = \omega \left(k \right)$ for small values of $k$ about the mentioned threefold solution
$\omega = 0$.
To this purpose we expand the dispersion relation (\ref{mo4}) about $\omega = 0$ according to
\begin{equation}
\omega = a _{L}k + b _{L}k ^{2} + .....
\label{mans0}
\end{equation}
and compare different orders of $k$ separately.
At lowest order which is $k ^{3}$ we find
\begin{eqnarray}
a _{L _{1}} &=& c _{s},
\label{mco1}\\
a _{L _{2}} &=& - c _{s},
\label{mco2}\\
a _{L _{3}} &=& 0.
\label{mco3}
\end{eqnarray}
The $k ^{4}$-order terms provide us with
\begin{eqnarray}
b _{L _{1,2}} &=& \frac{i}{2}
\left[\frac{\frac{4}{3}\eta + \zeta }{\rho + p}
+ \frac{\lambda T}{nT}\left(\frac{1}{c _{v}} - \frac{1}{c _{p}} \right)
+ \frac{1}{\rho + p}
\left(c _{s}^{2} - 2 \frac{\frac{\partial{p}}{\partial{T}}}
{\frac{\partial{\rho }}{\partial{T}}} \right)\right],
\label{mb12}\\
b _{L _{3}} &=& \frac{i \lambda }{n c _{p}},
\label{mb3}
\end{eqnarray}
where
\begin{equation}
c _{v} \equiv  \frac{1}{n}\frac{\partial{\rho }}{\partial{T}},
\mbox{\ \ \ \ \ \ }
c _{p} \equiv  c _{v}\frac{\rho + p}{n \frac{\partial{p}}{\partial{n}}}
c _{s}^{2}.
\label{mcvp}
\end{equation}
It is remarkable, that all the quantities $\beta _{0}$, $\beta _{1}$ and
$\beta _{2}$ cancel and do {\it not} influence the modes up to the order $k ^{2}$.
(This property is mentioned in \cite{HiLi2} following formula (40).
The modes (\ref{mans0}) with (\ref{mco1})-(\ref{mb3}) exactly coincide with those of the Eckart theory.
They were first derived by Weinberg \cite{Wein} and also in \cite{Gui}).
The coefficients (\ref{mb12}) describe a damping of the propagating
(with the sound velocities (\ref{mco1}) and (\ref{mco2})) modes, while the third mode, characterized by (\ref{mco3}) and (\ref{mb3}) is simply overdamped.
These modes characterize perturbations which are close to the perfect fluid behaviour. The dissipative terms describe small deviations from equilibrium.

\subsection{Transverse modes}
Using (\ref{ansq}) and (\ref{anspit}) we find
for the dissipative terms in (\ref{mtransv}) up to linear order
\begin{equation}
\epsilon _{amn}\dot{q}_{m,n}
= - \lambda T \left[2 \ddot{w}_{a}
+ \beta _{1}\epsilon _{amn}\ddot{q}_{m,n} \right]
\label{epsq}
\end{equation}
and
\begin{equation}
\epsilon _{amn}\pi _{m p,pn}
= - 2\eta \left[\Delta w _{a}
+ \beta _{2} \epsilon _{amn}
\dot{\pi}_{mp,pn}\right].
\label{epspi}
\end{equation}
For plane-wave solution of the type (\ref{mpw}), Eq. (\ref{mtransv}) becomes
\begin{equation}
2i \omega \left(\rho + p \right)w _{a}
- k _{p}k _{n}\epsilon _{amn}\pi _{mp}
+ i \omega \left(-ik _{n} \right)\epsilon _{amn}q _{m}
= 0,
\label{mtransv2}
\end{equation}
while the relations (\ref{epsq}) and (\ref{epspi}) transform into
\begin{equation}
i \omega \left(-ik _{n} \right)\epsilon _{amn}q _{m}
= 2 \lambda _{\beta }T \omega ^{2}w _{a}\
\label{epsq2}
\end{equation}
and
\begin{equation}
- k _{n}k _{b}\epsilon _{amn}\pi _{m b}
= 2 \eta _{\beta}k ^{2}w _{a},
\label{epspit2}
\end{equation}
respectively, where $\lambda _{\beta }$ is given by (\ref{lbeta}) and
$\eta _{\beta }$  by (\ref{etabeta}).
Use of (\ref{epsq2}) and (\ref{epspit2}) in (\ref{mtransv2}) provides us with the dispersion relation
\begin{equation}
i \omega \left(\rho + p \right) + \eta _{\beta }k ^{2}
+ \lambda _{\beta }T \omega ^{2} = 0,
\label{disptrans}
\end{equation}
which for $\eta _{\beta }\rightarrow \eta $ and
$\lambda _{\beta }\rightarrow \lambda $ coincides with the corresponding
relation of Eckart's theory.
We investigate the dispersion relation (\ref{disptrans}) analogously to its  counterpart (\ref{mdisp}) for longitudinal modes.
For $k \rightarrow 0$ we obtain, up to linear order in the dissipative terms,
\begin{equation}
i \omega \left[1 + i \omega
\frac{\lambda T}{\rho + p}\left(\beta _{1}\left(\rho + p \right)
-1 \right) \right]
\left[1 + 2 i \omega \eta \beta _{2} \right] = 0
\mbox{\ \ \ \ \ \ \ }
\left(k \rightarrow 0 \right).
\label{dispk0tr}
\end{equation}
For the Eckart theory ($\beta _{1}=\beta _{2}=0$) we obtain
$\omega = 0$ and
\begin{equation}
i \omega _{T _{E}} = \frac{\rho + p}{\lambda T},
\label{otrE}
\end{equation}
i.e., the same short-time instability as in the longitudinal case
(cf. Eq. (\ref{moE})).
For nonvanishing values of $\beta _{1}$ and $\beta _{2}$ we find,
except of the solution $\omega = 0$,
\begin{equation}
i \omega _{T _{1}} = - \frac{\rho + p}{\lambda T}
\frac{1}{\beta _{1}\left(\rho + p \right) - 1}
\label{otrI1}
\end{equation}
and
\begin{equation}
i \omega _{T _{2}} = - \frac{1}{2 \beta _{2}\eta }.
\label{otrI2}
\end{equation}
All the comments following Eq.~(\ref{mo2}) apply here also.
In particular, the instability of the Eckart theory is removed here as well for $\beta _{1}\left(\rho + p \right)>1$.
The modes (\ref{otrI1}) and (\ref{otrI2}) describe a damping on extremely short time-scales.
For comparison, (\ref{otrI1}) corresponds to (35) in \cite{HiLi2} while
(\ref{otrI2}) corresponds to (34) in \cite{HiLi2}.

The solutions $\omega = \omega \left(k \right)$ for small vales of $k$ about $\omega = 0$ are again found by an ansatz of the type (\ref{mans0}),
\begin{equation}
\omega = a _{T}k + b _{T}k ^{2} + ....
\label{otr}
\end{equation}
Using this ansatz in (\ref{disptrans}) and comparing terms of linear order in $k$ we find
\begin{equation}
a _{T} = 0,
\label{atr}
\end{equation}
while the order $k ^{2}$ yields
\begin{equation}
b _{T} = - i \frac{\eta }{\rho + p}.
\label{btr}
\end{equation}
This is exactly the nonpropagating, transverse shear mode of the first-order theory.
Consequently, neither the longitudinal nor the transverse modes are influenced by
the quantities $\beta _{0}$, $\beta _{1}$ and $\beta _{2}$ in linear and in quadratic orders in $k$.

\section{Conclusion}
With (\ref{mo0})-(\ref{mo2}) and (\ref{mans0})-(\ref{mb3}) for the longitudinal case,  as well as with (\ref{otrI1}) - (\ref{btr}) for the transverse one, we have obtained a comprehensive picture of the hydrodynamic modes in a dissipative fluid in the long-wavelength limit.
We have clarified the relationship between the mode structures of the MIS and
the Eckart theories.
None of the results here is new.
But I hope this specific pedagogically motivated presentation may be helpful for a better physical understanding and, possibly, may also be useful as a starting point and reference for further investigations in dissipative relativistic fluid dynamics.

\bibliography{bibliomode.bib}
\bibliographystyle{}

\end{document}